\documentclass[twocolumn,preprintnumbers,showpacs,showkeys,amsmath,amssymb,epsf]{revtex4}
\usepackage{graphicx}
\usepackage{dcolumn}
\usepackage{bm}
\usepackage{amsfonts}
\usepackage{amssymb}
\usepackage{bbm}
\usepackage{bbold}
\usepackage{color}
\begin{document}
\title{On Effective Stochastic Generators for Conditioned Dynamics at an Atypical Reaction-Diffusion Current }
\author{Pegah Torkaman}
\email{p.torkaman@basu.ac.ir}
\author{Farhad H. Jafarpour}
\email{farhad@ipm.ir}
\affiliation{Physics Department, Bu-Ali Sina University, 65174-4161 Hamedan, Iran}
\date{\today}
\begin{abstract}
We consider the fluctuations of a time-integrated particle current around an atypical value
in a generic stochastic Markov process involving classical particles with two-site interaction and hard-core repulsion on a 
finite one-dimensional lattice with open boundaries. We address the question of which interactions one has to impose on 
such process to make the atypical value of the current typical. It is known that a corresponding effective stochastic Markov 
process might exist whose typical value of the current is equal to the atypical value of the current in the original process 
within a time-translational invariant regime. This effective process has, in principle, non-local transition rates. Nevertheless, it 
turns out that under some conditions the stochastic generator of the effective process has the same dynamical rules as 
the stochastic generator of the original process. We find these conditions and show that our approach can be generalized to 
any time-integrated observable.
\end{abstract}
\pacs{05.40.-a,05.70.Ln,05.20.-y}
\keywords{non-equilibrium systems, stochastic particle dynamics (theory), effective dynamics, current fluctuations, large deviations}
\maketitle
\section{Introduction}
\label{I}
Rare events and their characterizations are of vital importance in different contexts of physics. 
These phenomena take place on a timescale much larger than the timescales characterizing 
the microscopic dynamics of the system. For example nucleation of crystals relies on a rare 
event i.e. the formation of the critical nucleus~\cite{AF01}. Protein-folding is also a rare event. In spite of 
an astronomical number of possible configurations for a protein, it folds into a unique native 
conformation~\cite{PF}. Another classical example includes phase transformation for which the dynamics 
might be governed by rare events~\cite{WWV05}. 

In a general stochastic process the effective interactions that induce particular rare events are generally  
very complicated. In an equilibrium stochastic process the principle of detailed balance requires 
that the transition rates between a pair of microstates in the canonical ensemble of the process 
satisfy certain relation i.e. the ratio of rates for a transition and its time-reverse is given by the Boltzmann factor.  
As a sub-set of this equilibrium ensemble one can consider a particular driven ensemble consisting 
of phase-space paths (sometimes called an ensemble of trajectories) for which the mean flux of an 
observable on those paths is fixed. The existence of a net flux implies that the we are dealing with 
a constraint driven dynamics. Using the Bayes' theorem, it has been shown that the transition rates 
of this driven stochastic process with a given flux are related to those of the equilibrium system~\cite{E04,E05}. 
Although unphysical transitions in the original equilibrium system, which might violate the relevant 
physical laws, will not appear in the driven dynamics; however, the transition rates of the driven 
system might be non-local. The non-equilibrium counterpart to the equilibrium detailed balance derived 
in~\cite{E04,E05} results in a set of invariant quantities in the driven system analogues to the equilibrium one. 
This provides us with exact relations which help us calculate the transition rates in the driven 
system~\cite{BE08}. The results obtained in~\cite{E04,E05,BE08} can be reproduced  by maximizing the dynamical 
entropy in the presence of appropriate constraints~\cite{M11}.

It is long known that in order to study the dynamics of a stochastic process conditioned on atypical values of a 
time-integrated observable in the steady state of a generic stochastic process system, whether this observable 
depends on microstates or transitions between a pair of microstates,  one can use the concept of biased ensemble 
of trajectories~\cite{LAV07}-\cite{PSS10}. This can be done by introducing a biasing field conjugated to the mean 
value of the observable. During a long observation time $t$ the ensemble average of a given observable in this 
biased ensemble of trajectories might depend on time and therefore, the time-translation invariance might be broken. However,  
there exists a time interval $[t_1,t_2]$, with $t_1$ and $t_2$ being far from the initial time and the final time $0$ and $t$ 
respectively, where the time-translation invariance is held i.e. the ensemble average of the observable under investigation in this 
time interval is independent of time. It has been shown that being in the steady state and during this 
time-translational invariant regime $[t_1,t_2]$ the biased trajectories of the original process coincide with unbiased trajectories
of an effective (or auxiliary) stochastic process~\cite{JS10}. Hence the average of the observable over the steady-state distribution of the 
effective stochastic process will be equal to its average over the biased ensemble of trajectories during the time-translational invariant regime. 
The effective stochastic process is a conditioning-free process describing the problem of conditioning a Markov process on 
an atypical value of the dynamical observable. The mathematical relation between these processes is given by a generalization 
of Doob's h-transform~\cite{CT1}. It was shown that the effective process can be represented as a process satisfying various 
variational principles or a control process optimizing functionals related to the large deviations of the conditioning dynamical 
observable~\cite{CT2}. The connection between effective interactions and the theory of optimal control has also been studied 
in~\cite{JS15}. The analysis of the effective interaction in this way is used, for example, in the East model as one of the kinetically 
constrained models consisting of interacting spins in Glass-forming systems~\cite{JS14}. 

The effective stochastic process consists of those interactions one has to impose on the original stochastic process to make
atypical behavior typical. As in the equilibrium case explained above, the effective process might be unphysical in the sense that its transition rates 
might be non-local~\cite{JS10}. This means that the original stochastic process and its corresponding effective stochastic process might not share similar
features such as the range of interactions. The one-dimensional classical Ising chain, which exhibits ferromagnetic ordering in its biased
ensemble of trajectories, is an example which reveals this feature~\cite{JS10}. Similar examples are studied in~\cite{JS14,PS11}. 

A natural question that might arise is that under what conditions the corresponding effective stochastic process of a stochastic process
with conditioned dynamics is physical in the sense that, in comparison with the original dynamics, no non-local transitions appear in 
the effective dynamics. In other words, under what conditions imposed on the microscopic reaction rates or for which atypical values of the 
observable, the stochastic generators of these two processes are exactly the same (up to a rescaling of the microscopic
dynamical rules). This might not be valid for all atypical values of the observable; however, as we will see one might be able to find at least an 
atypical value of the observable for which the dynamics of the original stochastic process and its corresponding effective process
share the same features at that point. 

In present paper we are going to address the above question for a specific class of stochastic Markov processes consisting of interacting classical particles 
on a one-dimensional lattice with open boundaries. We assume that the particles are subjected to nearest-neighbor interactions in the bulk
of the lattice while they can enter or leave the lattice from both the first and the last lattice sites. Considering the total reaction-diffusion current as
a physical observable, we require that the corresponding effective stochastic process consists of exactly the same interactions 
in the bulk and at the boundaries of the lattice. In other words, we require that, up to a rescaling of the transition rates, the 
stochastic generator of the effective stochastic process is exactly the same as the stochastic generator of the original stochastic 
process conditioned on an atypical value of the total particle current. We show that, given that there are some constraints on the 
dynamical rules of the original stochastic process, there is at most a single atypical value of the average current at which this property 
might be held. 

A couple of examples are given in the present paper. In the first example the Asymmetric Simple Exclusion Process (ASEP) is 
considered on an open lattice. In this system the particles with hard-core interactions perform a continuous-time simple random 
walk on an open lattice with the possibility of entering or leaving the lattice from both the first and the last lattice sites. Considering 
a barrier-free hopping of particles between the bulk of the lattice and the particle reservoirs with the same hopping rates as inside 
the bulk and assuming that the diffusion rates are biased to the right, it turns out that the conditions under which the effective and 
the original ASEP share the same features, restrict us to an atypical value of the particle current which is lower than the average 
particle current in the steady state of the original ASEP. On the other hand, it can be seen that the effective dynamics is exactly 
the one for the ASEP but with a reversed driving force (i.e. the diffusion rates are biased to the left). This phenomenon has already 
been observed in a recent work~\cite{BS13}. It has been shown that under some constraints the steady state of the effective ASEP 
can be written as a superposition of antishocks. 

In the second example we consider an Asymmetric Kawasaki-Glauber Process (AKGP) on a one-dimensional lattice with open 
boundaries~\cite{KJS03}. In this case the non-zero rates are the death and branching rates as well as the hopping rate to the left. It is known 
that stable shocks can develop in the AKGP. Hence, a linear superposition of them can be used to construct 
its steady-state which consists of a hight-density phase and a low-density phase. As we will see by fine tuning the microscopic 
reaction rates the stochastic generator of the effective process can be brought to the form of the stochastic generator of the AKGP. 
Being in either of the static phases, the atypical current at which this phenomenon takes place can be lower or higher than the typical
value of the current in the steady state, depending on the microscopic reaction rates. 

Finally we will bring the third example in which the above mentioned phenomenon can happen for an atypical value of a 
non-entropic particle current. While the current in the ASEP is entropic and for the AKGP is zero (since the steady state 
is an equilibrium one), interestingly the large deviation function for the current in our third example satisfies the 
Gallavotti-Cohen-like symmetry~\cite{BCHM12}. On the other hand, the atypical value of the current at which the above mentioned 
phenomenon happens, is exactly equal to the typical value of the current in the steady state but with the opposite sign. 

This paper is organized as follows. In section II we start with mathematical preliminaries and tools.  In section III we define the 
reaction-diffusion current and find the conditions under which the stochastic generator of the effective process is equivalent with the 
stochastic generator of the original process. In section IV we will bring three examples to show how our constraints determine the 
effective dynamics. The generalization is brought in section V. The last section is devoted to the outlook and conclusion. 

\section{Mathematical Tools: A short review}
\label{II}
We start with a stochastic Markov process in continuous-time. This is defined through a set of configurations denoted by 
$ \{C\}$ and stochastic transition rates $\omega_{C \to C'}$ between these configurations. Considering the complete basis vector $\{ | C \rangle \}$, the 
probability of finding the system in configuration $C$ at time $t$ is given by $P(C,t)=\langle C | P(t) \rangle$ where the ket  $| P(t) \rangle$ evolves in time 
according to the following master equation~\cite{SCH01}
\begin{equation}
\label{ME}
\frac{d}{dt} | P(t) \rangle = \hat{\cal{H}} | P(t)\rangle
\end{equation}
in which the stochastic generator or Hamiltonian $\hat{\cal{H}}$ is a square matrix with the following matrix elements
$$
\langle C | \hat{\cal{H}} | C' \rangle = \omega_{C' \to C} - \delta_{C,C'} \sum_{C'' \neq C} \omega_{C \to C''} \; .
$$
Let us now consider a reaction-diffusion system consisting of interacting classical particles on a one-dimensional lattice of length $L$
which is modeled by a stochastic Markov process in continuous-time. Being in the steady state, we denote the mean 
(or typical) value of the reaction-diffusion current as $J^{\ast}$. Let ${\cal J}$ be the number of reaction
and diffusion processes which contribute to the total reaction-diffusion current of the system up to the time $t$. This quantity is extensive 
with respect to $t$ and $L$. For a finite $L$ we define the space-time average of the total reaction-diffusion current as $J={\cal J}/(Lt)$ which is 
a time dependent quantity. During a long-time interval $t$ the probability to observe an atypical mean $J\neq J^{\ast}$ is exponentially 
small in $L$ and $t$. The large deviation property requires $P({\cal J})\propto \exp(-I(J)Lt)$ where $I(J)$ is called the rate function. 
Now $\lim_{t \to \infty} \ln \langle e^{-s{\cal J}}\rangle /(Lt)$ gives the cumulant generating function of the current $J$
in which $\langle e^{-s{\cal J}}\rangle=\sum_{\cal J}e^{-s{\cal J}} P({\cal J})$ and that $s$ is called the counting field 
conjugated to the mean current $J$~\cite{BS13,BS2013}.  

We aim to study the dynamics of the above mentioned system conditioned on an atypical value of  the current $J$.
We define ${\cal J}_{C \to C'}$ as an increment for this current during transition from configuration $C$ to $C'$.
It is known that the generating function of ${\cal J}$ defined above is given by 
$\langle e^{-s {\cal J}} \rangle = \langle \mathbb{1} | P_s(t) \rangle$ where $ \langle \mathbb{1} |=\sum_{C} \langle C |$ is 
called the summation vector,  and that $| P_s(t) \rangle$ should be obtained from the following master equation~\cite{T09}   
\begin{equation}
\label{SME}
\frac{d}{dt} | P_s(t) \rangle = \hat{\cal{H}}(s) | P_s(t)\rangle \; .
\end{equation}
The operator $\hat{\cal{H}}(s)$ in~(\ref{SME}) is non-stochastic and called the modified Hamiltonian of the system with the 
following matrix elements
$$
\langle C | \hat{\cal{H}}(s) | C' \rangle =e^{-s {\cal J}_{C' \to C}} \omega_{C' \to C} - \delta_{C,C'} \sum_{C'' \neq C} \omega_{C \to C''} \; .
$$
The counting field $s$ can be interpreted as a biasing field in the ensemble 
of dynamical trajectories which is sometimes called the $s$-ensemble. The role of $s$ in the dynamical ensemble is similar to the parameter 
$\beta$ (inverse of temperature) in the conventional equilibrium canonical ensemble. Using this biased ensemble one can study the dynamics 
of system during the observation time $t$ conditioned on a given value of the mean current $J$. Fixing some $s\neq 0$ correspond to studying 
those realizations of the process in which $J$ fluctuates around some atypical mean value~\cite{JS10}. This approach is sometimes called 
the grand canonical conditioning which corresponds to constructing a canonical ensemble of trajectories~\cite{BS13,BS2013,PSS10}. 
According to our notation the positive (negative) values of the counting field $s$ 
correspond to the atypical values of the current lower (higher) than the typical value of the current in the steady state. 
The time-evolution generator or modified Hamiltonian for the conditioned dynamics $\hat{\cal{H}}(s)$ is a non-stochastic operator  
which does not conserve probability. The sum of unnormalized probabilities is called the dynamical partition function of this dynamical 
ensemble and is given by $Z(s,t)=\langle \mathbb{1} |P_s(t) \rangle$. The logarithm of this quantity plays the role of the dynamical free 
energy of system which determines its dynamical phase behavior~\cite{LAV07}. 

Following the discussion in section~\ref{I}, there is a time-translational invariant regime during which one can construct  
an effective (or auxiliary) stochastic process whose unbiased dynamics produces the same value of mean current as the 
conditioned (or biased) dynamics explained above does during that time interval~\cite{JS10}. Considering the eigenvalue 
equations for the modified Hamiltonian $\hat{\cal{H}}(s)$
$$
\begin{array}{l}
\hat{\cal{H}}(s) | \Lambda(s) \rangle = \Lambda (s) | \Lambda(s) \rangle \; , \\
\hat{\cal{H}}(s) \langle \tilde{\Lambda}(s) | = \Lambda (s) \langle \tilde{\Lambda}(s) | 
\end{array}
$$
it has been shown that the stochastic generator of this effective stochastic process is given by~{\cite{JS10}}
\begin{equation}
\label{relation}
\hat{\cal{H}}_{eff}(s) =\hat{U} \hat{\cal{H}}(s) \hat{U}^{-1}-\Lambda^{{\ast}}(s) 
\end{equation}
in which ${\hat U}$ is a diagonal matrix with the matrix element $ \langle C | \hat{U} | C \rangle = \langle \tilde{\Lambda}^{{\ast}}(s) |C \rangle $ 
and the asterisk stands for the largest eigenvalue and corresponding left and right eigenvectors of $\hat{\cal H}(s)$. 
The off-diagonal matrix elements of the operator $\hat{\cal{H}}_{eff}(s)$ in~(\ref{relation}) are given by 
 \begin{equation}
\label{element}
\langle C | \hat{\cal{H}}_{eff}(s) | C' \rangle =\frac{
 \langle \tilde{\Lambda}^{{\ast}}(s) |C \rangle
\langle C | \hat{\cal{H}}(s) | C' \rangle}
{ \langle \tilde{\Lambda}^{{\ast}}(s) |C' \rangle} \; .
\end{equation}
It is easy to see that for the systems with a finite-dimensional configuration space $\Lambda^{{\ast}}(s)=\lim_{t \to \infty} \ln \langle e^{-s{\cal J}}\rangle /t$~\cite{T09}.

\section{Equivalence of original and effective dynamics}
\label{III}
In this section we limit ourselves to a family of single-species reaction-diffusion systems of classical particles with nearest-neighbor interactions
in the bulk of a one-dimensional lattice with open boundaries from there the particles can enter or leave the lattice. We aim to find
the conditions under which the effective Hamiltonian $\hat{\cal H}_{eff}(s)$ of this family is similar to that of the original process 
conditioned on some atypical mean current $J$ during its time-translational invariant regime, in the sense that the effective Hamiltonian 
consists of exactly the same type of interactions in the bulk and boundaries of the lattice. The simplest choice is where $\hat{U}$ in~(\ref{relation}) is an 
identity matrix. This means that the modified Hamiltonian $\hat{\cal{H}}(s)$ and the effective Hamiltonian $\hat{\cal{H}}_{eff}(s)$ differ 
from each other by a constant which, according to~(\ref{relation}), is the largest eigenvalue of the $\hat{\cal{H}}(s)$. 

For the above mentioned family of stochastic processes the Hamiltonian $\hat{\cal{H}}$ can be written as
\begin{equation}
\begin{array}{lll}
\label{hamiltonian}
\hat{\cal{H}} & = & \hat{\cal{L}} \otimes {\cal I}^{\otimes (L-1)} \\ \\ 
   & + &  \sum_{k=1}^{L-1} \big ( {\cal I}^{\otimes (k-1)}\otimes  \hat{h}  \otimes {\cal I}^{\otimes (L-k-1)}\big )\\  \\
   & + &  {\cal I}^{\otimes (L-1)}\otimes \hat{\cal{R}}
\end{array}
\end{equation}
in which $\cal I$ is a $2 \times 2$ identity matrix. Introducing the basis kets
$$
\vert  \emptyset \rangle = \left( \begin{array}{c}
1\\
0\\
\end{array} \right)\, ,\;\;  
\vert A \rangle=\left( \begin{array}{c}
0\\
1\\
\end{array} \right)\,
$$
in which $\emptyset$ and $A$ correspond to a vacancy and an occupied lattice site respectively, 
the matrix representation of $\hat{h}$ in the basis of $\{ \emptyset \emptyset, \emptyset A, A \emptyset, A A\}$ 
and that of $\hat{\cal{L}}$ and $\hat{\cal{R}}$ in the basis of $\{ \emptyset,A \}$ are given by
$$
\begin{array}{c}
\hat{h}=\left( \begin{array}{cccc}
\omega_{11} & \omega_{12}&\omega_{13}& \omega_{14}\\
\omega_{21} & \omega_{22}&\omega_{23}& \omega_{24}\\
\omega_{31} & \omega_{32}&\omega_{33}& \omega_{34}\\
\omega_{41} & \omega_{42}&\omega_{43}& \omega_{44}\\
\end{array} \right)\, , \\ \\
\hat{\cal{L}}=\left( \begin{array}{cc}
-\alpha & \gamma\\
\alpha & -\gamma\\
\end{array} \right)\, , 
\hat{\cal{R}}=\left( \begin{array}{cc}
-\delta & \beta\\
\delta & -\beta\\
\end{array} \right)\, . 
\end{array}
$$
The diagonal elements of $\hat{h}$ are given by $\omega_{ii}=-\sum_{j \neq i} \omega_{ji}$. As can be seen the parameters 
$\alpha$ and $\gamma$ ($\delta$ and $\beta$) are the injection and extraction rates of particles for the left (right) boundary
respectively. 

Let us consider the total reaction-diffusion current as the proper dynamical observable. The time-derivative of 
the average local density of particles is related to the average particle current through the following continuity equation 
\begin{equation}
\label{ce}
\frac{d}{dt} \langle \rho_k \rangle =\langle j_{k-1} \rangle-\langle j_{k} \rangle+S_{k} \;\; \mbox{for} \;\; 1 \le k \le L
\end{equation}
in which $\langle j_{k} \rangle$ is defined as the average local particle current from the lattice site $k$ to $k+1$
and is given by 
\begin{eqnarray}
\label{jk}
\langle j_{k} \rangle  &=& [(\omega_{21}-\omega_{31}) \langle (1-\rho_{k})(1-\rho_{k+1}) \rangle \nonumber \\
&-&(\omega_{12}+\omega_{42}+\omega_{32}) \langle (1-\rho_{k})\rho_{k+1} \rangle \nonumber \\
&+&(\omega_{43}+\omega_{13}+\omega_{23}) \langle \rho_{k}(1-\rho_{k+1}) \rangle \\ 
&+&(\omega_{24}-\omega_{34})  \langle \rho_{k} \rho_{k+1} \rangle] (1-\delta_{k,L}) (1-\delta_{k,0}) \nonumber  \\
&+& (\beta \langle \rho_{k}\rangle -\delta \langle 1-\rho_{k}\rangle)\delta_{k,L}  \nonumber  \\ 
&+& (\alpha \langle 1-\rho_{k+1}\rangle -\gamma \langle \rho_{k+1}\rangle)\delta_{k,0} \nonumber
\end{eqnarray}
for $k=0,\cdots,L$. $S_{k}$ is the source term. For the details of derivation~(\ref{jk}) see Appendix. 
The average total reaction-diffusion current, which includes the contribution of all bonds of the lattice, is now given by
\begin{equation}
\label{RDC}
\langle J \rangle=\frac{1}{L}\sum_{k=0}^{L} \langle j_{k} \rangle  \; .
\end{equation}
Considering the total reaction-diffusion current defined in~(\ref{RDC}) as a dynamical observable, the modified 
Hamiltonian $\hat{\cal{H}}(s)$  is given by
\begin{equation}
\label{hs}
\begin{array}{lll}
\hat{\cal{H}}(s) & = & \hat{\cal{L}}(s) \otimes {\cal I}^{\otimes (L-1)} \\ \\
& + &  \sum_{k=1}^{L-1} \big ( {\cal I}^{\otimes (k-1)}\otimes  \hat{h}(s)  \otimes {\cal I}^{\otimes (L-k-1)}\big )\\ \\
& + & {\cal I}^{\otimes (L-1)}\otimes \hat{\cal{R}}(s)
\end{array}
\end{equation}
in which
$$
\begin{array}{c}
\hat{h}(s)=\left( \begin{array}{cccc}
\omega_{11} & \omega_{12}e^{s}&\omega_{13}e^{-s}& \omega_{14}\\
\omega_{21}e^{-s} & \omega_{22}&\omega_{23}e^{-s}& \omega_{24}e^{-s}\\
\omega_{31}e^{s} & \omega_{32}e^{s}&\omega_{33}& \omega_{34}e^{s}\\
\omega_{41} & \omega_{42}e^{s}&\omega_{43}e^{-s}& \omega_{44}\\
\end{array} \right)\, , \\ \\
\hat{\cal{L}}(s)=\left( \begin{array}{cc}
-\alpha & \gamma e^{s}\\
\alpha e^{-s} & -\gamma\\
\end{array} \right)\, , 
\hat{\cal{R}}(s)=\left( \begin{array}{cc}
-\delta & \beta e^{-s}\\
\delta e^{s} & -\beta\\
\end{array} \right)\, . 
\end{array}
$$
The increment of the current for each reaction process or diffusion process can be understood from~(\ref{jk}).
Fixing the counting field $s$, corresponding to study of an atypical value of the current ${\cal J}$, and trying to find the 
effective Hamiltonian $\hat{\cal H}_{eff}(s)$ can be a formidable task.  

Considering~(\ref{element}) one should note that $\hat{\cal{H}}_{eff}(s)$ can not necessarily be written in the two-site interaction form 
though $\hat{\cal{H}}$ is of the form~(\ref{hamiltonian}).  Generally speaking, a simple system might have complex effective interactions. 
As a matter of fact, it has been shown that even for a system with short-range interactions the effective interactions might be 
long-range~\cite{JS10,JS14,PS11}.  However, as we will see, there might be a value of $s=s_0$ at which the stochastic Hamiltonian 
of effective dynamics is similar to~(\ref{hamiltonian}) which means it involves nearest-neighbor interactions in the bulk and single-site 
interactions with the reservoirs at the boundaries. 

We have found that under the following constraints 
\begin{eqnarray}
\label{constraint}
e^{s_0} &=& \frac{\omega_{13}+\omega_{23}+\omega_{43}-2\omega_{21}}{\omega_{12}+\omega_{32}+\omega_{42}-2\omega_{31}}  \nonumber \\ 
 &=&  \frac{\omega_{13}+\omega_{23}+\omega_{43}-\omega_{21}-\alpha}{\gamma-\omega_{31}}   \nonumber    \\ 
 &= & \frac{\beta-\omega_{21}}{\omega_{12}+\omega_{32}+\omega_{42}-\omega_{31}-\delta}   \\ 
 &= & \frac{\omega_{24}-\omega_{21}}{\omega_{34}-\omega_{31}} >0 \nonumber
\end{eqnarray}
the effective Hamiltonian $\hat{\cal {H}}_{eff}(s)$ has the form of~(\ref{hamiltonian}) with
$$
\begin{array}{l}
\hat{h}_{eff} (s_0)=\left( \begin{array}{cccc}
\omega_{11}' & \omega_{12}e^{s_0}&\omega_{13}e^{-s_0}& \omega_{14}\\
\omega_{21}e^{-s_0} & \omega_{22}'&\omega_{23}e^{-s_0}& \omega_{24}e^{-s_0}\\
\omega_{31}e^{s_0} & \omega_{32}e^{s_0}&\omega_{33}'& \omega_{34}e^{s_0}\\
\omega_{41} & \omega_{42}e^{s_0}&\omega_{43}e^{-s_0}& \omega_{44}'\\
\end{array} \right)\, , \\ \\
\hat{\cal{L}}_{eff}(s_0) =\left( \begin{array}{cc}
-\alpha e^{-s_0}& \gamma e^{s_0}\\
\alpha e^{-s_0} & -\gamma e^{s_0}\\
\end{array} \right)\, , \\ \\
\hat{\cal{R}}_{eff} (s_0)=\left( \begin{array}{cc}
-\delta e^{s_0} & \beta e^{-s_0}\\
\delta e^{s_0} & -\beta e^{-s_0}\\
\end{array} \right)
\end{array}
$$
where the diagonal elements of $\hat{h}_{eff} $ are given by $\omega_{ii}'=-\sum_{j \neq i} ({\hat h}_{eff}(s_0))_{ji}$ which is the 
stochasticity condition for the effective Hamiltonian. The largest eigenvalue of $\hat{\cal{H}}(s)$ at $s=s_0$ turns out to be
\begin{equation}
\begin{array}{lll}
\label{eigenvalue}
\Lambda^{\ast}(s_0)&=&\Big(\alpha+(L-1)\omega_{21}\Big) (e^{-s_0}-1)\\ \\
        &+&\Big(\delta+(L-1)\omega_{31}\Big) (e^{s_0}-1) 
\end{array}
\end{equation}
where its corresponding left eigenvector is given by $ \langle \tilde{\Lambda}(s_0) | =  \langle \mathbb{1} | $.
One should note that the above left eigenvector results in the following exact expression for the generating function 
of the current at $s=s_0$
\begin{eqnarray}
 \langle e^{-s_0 J} \rangle &=& \langle \mathbb{1} | P_{s_0}(t) \rangle \nonumber \\
 &=&\langle \mathbb{1} |e^{t \hat{\cal{H}}_{s_0}} |P_{s_0}(0) \rangle \\
 &=& e^{t\Lambda^*(s_0)}  \langle \mathbb{1} | P_{s_0}(0) \rangle  \nonumber\\
 &=& e^{t\Lambda^*(s_0)} \nonumber \; .
\end{eqnarray}
Depending on the process under investigation, the eigenvalue~(\ref{eigenvalue}) might depend linearly on the system size $L$. 
There are two cases for which the largest eigenvalue $\Lambda^*(s_0)$ can be independent of the system size. The first case is where 
$\omega_{21}=\omega_{31}=0$ while the rest of the reaction rules satisfy~(\ref{constraint}). In the second case $s_0=\ln \frac{\omega_{21}}{\omega_{31}}$ while
\begin{eqnarray}
e^{s_0} &=& \frac{\omega_{24}}{\omega_{34}} = \frac{\omega_{13}+\omega_{23}+\omega_{43}}{\omega_{12}+\omega_{32}+\omega_{42}}  \nonumber \\ 
 &=&  \frac{\omega_{13}+\omega_{23}+\omega_{43}-\alpha}{\gamma}   \nonumber    \\ 
 &= & \frac{\beta}{\omega_{12}+\omega_{32}+\omega_{42}-\delta}  >0 \; .
\end{eqnarray}

\section{Examples}
\label{IV}
In this section a couple of examples are presented to show how the conditions obtained in the previous section might generate 
interesting results. In the first example we consider the ASEP with open boundaries as explained in the introduction. In the bulk 
of the lattice the particles hop to the right and left according to the following rules: 
\begin{equation}
\label{aseprules}
\begin{array}{ll}
A \; \emptyset \; \longrightarrow \; \emptyset \; A \quad \mbox{with the rate} \quad \omega_{23}=p \; ,\\
\emptyset\; A  \; \longrightarrow \; A \; \emptyset \quad \mbox{with the rate} \quad \omega_{32}=q
\end{array}
\end{equation}
All other reaction rates in the bulk of the lattice are zero. 
The particles are also injected and extracted from the boundaries of the lattice with the rates $\alpha,\; \gamma,\; \beta$ and $\delta$ as 
explained in the previous section. The constraints~(\ref{constraint}) give
\begin{eqnarray}
\label{asep1}
e^{s_0} =\frac{p}{q} , \;\; \frac{\alpha}{p}+\frac{\gamma}{q}=\frac{\beta}{p}+\frac{\delta}{q}=1 \; .
\end{eqnarray}
Let us assume that the density of the particles at the left and right boundaries is kept fixed, using two particle reservoirs,
at the values  $\rho_1$ and $\rho_2$ respectively. This can be done by choosing barrier-free boundary rates defined as~\cite{BS13}
$$
\alpha=p \rho_1, \;\; \gamma=q (1-\rho_1) , \;\; \beta=p (1-\rho_2), \;\; \delta=q \rho_2 \; .
$$
In this case the only constraint which remains will be
\begin{equation}
\label{asep2}
e^{s_0} =\frac{p}{q} 
\end{equation}
and the eigenvalue is given by 
\begin{equation}
\Lambda^{\ast}(s_0)= -(p-q)(\rho_1-\rho_2)\; .
\end{equation}
By substituting $s_0$ in $\hat{\cal H}(s)$ it is easy to see that the effective Hamiltonian can be obtained from the Hamiltonian of the 
original ASEP by exchanging $p$ and $q$ (or reversal $p \leftrightarrow q$ of the particle hopping rates). This observation has an 
interesting consequence. Given that one chooses barrier-free boundary rates, it is known that the steady state of the ASEP 
can be written as a linear superposition of Bernoulli measures with a step-function structure provided that the following constraint is 
satisfied~\cite{KJS03} 
$$
\frac{\rho_2 (1-\rho_1)}{\rho_1 (1-\rho_2)} = \frac{p}{q} \;.
$$
Hence, if one chooses $p>q$ then the constraint requires $\rho_1 < \rho_2$ and this is what we call a shock structure. 
Now, since the effective Hamiltonian is exactly the same as the original Hamiltonian but with reversed hopping rates
one can conclude that the steady state of the effective Hamiltonian can also be written as a superposition of Bernoulli 
measures with a step-function structure. In this case one should have
$$
\frac{\rho_2 (1-\rho_1)}{\rho_1 (1-\rho_2)} = \frac{q}{p} 
$$  
and since $p>q$ then $\rho_1 > \rho_2$. In comparison to the definition of a shock structure this is called an antishock.  
One should note that since for $p>q$ we have $s_0>0$, then atypical value of the current is always lower than the typical 
value in the steady state. This has already been observed and discussed with more detail in~\cite{BS13}. 

In the second example we consider an asymmetric Kawasaki-Glauber process which contains the following
reaction rules in the bulk of the lattice:
\begin{equation}
\label{akgprules}
\begin{array}{ll}
\emptyset\; A  \; \longrightarrow  \;   \emptyset \; \emptyset   \quad & \mbox{with the rate} \quad \omega_{12} \; ,\\
\emptyset\; A  \; \longrightarrow  \;   A \;               \emptyset  \quad & \mbox{with the rate} \quad \omega_{32} \; ,\\
\emptyset\; A  \; \longrightarrow  \;   A \; A                              \quad & \mbox{with the rate} \quad \omega_{42} \; ,\\
A \; \emptyset \; \longrightarrow \; \emptyset \;  \emptyset    \quad & \mbox{with the rate} \quad \omega_{13} \; ,\\
A \; \emptyset \; \longrightarrow \; A \; A                                \quad & \mbox{with the rate} \quad \omega_{43} \; .
\end{array}
\end{equation}
The only non-zero boundary rates $\alpha$ and $\beta$ define the injection and extraction of the particles at the left and right 
boundaries of the lattice respectively. The constraints (\ref{constraint}) for this process are
$$
e^{s_0} =\frac{\omega_{13}+\omega_{43}}{\omega_{12}+\omega_{42}+\omega_{32}}\;, \;\;\alpha=\beta=\omega_{13}+\omega_{43} \; .
$$
It is known that the steady state of this process, without any constraints on the microscopic reaction rates, can be written as a superposition of 
stable Bernoulli shock measures~\cite{KJS03}. It has also been shown that the microscopic position of each shock performs a biased random 
walk on the lattice. Now, following our discussion in the first example, we conclude that the steady state of the effective dynamics can be 
written in terms of superposition of Bernoulli shock measures (and not antishocks). Note that in the steady state 
of the original process the system undergoes a static phase transition between a low-density and a high-density phase depending on the
values of $\omega_{13}$ and $\omega_{43}$. It is worth mentioning that, being in either of these static phases, the dynamics 
can be either conditioned on a lower than typical or a higher than typical value of the total average current.  

\begin{figure}
\includegraphics[width=80mm]{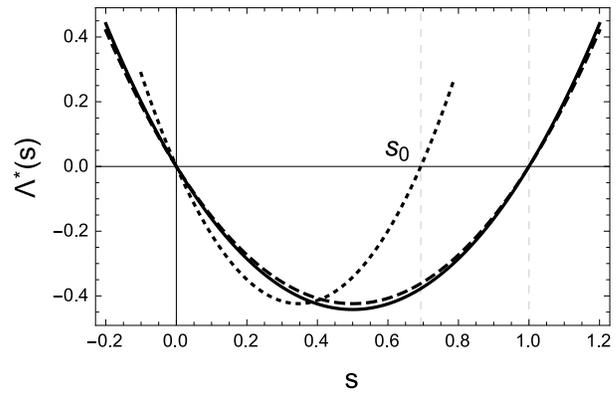}
\caption{\label{fig1} The plot of the numerically calculated largest eigenvalue of the modified Hamiltonian.
The dotted curve is $\Lambda^*_{cur} (s)$ for $\omega_{12} = 1, \omega_{13} = 2, \omega_{21} = 0.8$ for a lattice of length $L=6$. 
The solid line and the dashed line correspond to $\Lambda^*_{ent} (s)$ and $\Lambda^*_{cur} (Es)$ respectively. The vertical line is 
$s_0=\ln(\omega_{13}/\omega_{12})=0.69$. See inside the text for more information.}
\end{figure}

The time-integrated currents are generally either entropic which satisfy the Gallavotti-Cohen symmetry such as the one studied in~\cite{TJ13} 
or non-entropic which satisfy the Gallavotti-Cohen-like symmetry such as the one studied in~\cite{BCHM12}.
In the third example we introduce a non-entropic reaction-diffusion current which satisfies the Gallavotti-Cohen-like symmetry with the 
mentioned property. It turns out that under some constraints the value of the conjugated field $s_0$ can be 
located on the symmetry point corresponding to $s=0$ which means we have $\Lambda^{\ast}(0)=\Lambda^{\ast}(s_0\neq 0)=0$. This 
indicates that the absolute values of the atypical current and the typical current are equal; however, they flow in opposite directions.  

Our third example consists of birth and death processes in the bulk of the lattice with the rates ($\omega_{21},\omega_{31}$) and 
($\omega_{12},\omega_{13}$) respectively which can be demonstrated as follows:
\begin{equation}
\label{bdrules}
\begin{array}{ll}
\emptyset\; \emptyset  \; \longrightarrow  \;  \emptyset \; A  \quad & \mbox{with the rate} \quad \omega_{21} \; ,\\
\emptyset\; \emptyset  \; \longrightarrow  \; A \; \emptyset   \quad & \mbox{with the rate} \quad \omega_{31} \; ,\\
\emptyset\; A  \; \longrightarrow  \;   \emptyset \;  \emptyset    \quad & \mbox{with the rate} \quad \omega_{12} \; ,\\
A \; \emptyset \; \longrightarrow \; \emptyset \;  \emptyset    \quad & \mbox{with the rate} \quad \omega_{13} \; .
\end{array}
\end{equation}
The particles are allowed to enter or leave the lattice from both boundaries; however, 
the boundary rates $\alpha$ and $\delta$ are assumed to satisfy the following constraints
$$\alpha=\omega_{21},\; \delta=\omega_{31}\; .$$ 
Now the constraints~ (\ref{constraint}) lead us to 
\begin{eqnarray*}
&& s_0=\ln \frac{\omega_{21}}{\omega_{31}}=\ln\frac{\omega_{13}}{\omega_{12}} , \\
&& \gamma=\omega_{12}-\omega_{31},\\
&& \beta=\omega_{13}-\omega_{21} \;.
\end{eqnarray*}
The largest eigenvalue of the modified Hamiltonian for the entropy production $\Lambda^*_{ent}(s)$ and that of the total diffusion-reaction 
current $\Lambda^*_{cur} (s)$ are numerically calculated and plotted in FIG.\ref{fig1}. We have also plotted $\Lambda^*_{cur} (E s)$ where $E=s_0$.
As can be seen $\Lambda^*_{ent}(s)$ does not lie on $\Lambda^*_{cur} (E s)$ and therefore the total diffusion-reaction current is non-entropic ~\cite{BCHM12}. 

\section{generalizations}
\label{V}
The above discussion can be generalized to any arbitrary time-integrated observable (which is not necessarily the particle current) in 
a continuous-time stochastic Markov process with a stochastic generator of type~(\ref{hamiltonian}) and a finite configuration space. 
These observables can be fluxes or currents which depend on the transitions between configurations or microstates, such as the one 
we explained in this paper. Alternatively we can consider those time-integrated observables that might have merely a spatial nature 
such as dynamical activity~\cite{LAV07} or energy~\cite{JS10}. 

For the fluxes or currents which are defined on the basis of transitions between configurations, we consider the 
increment $\theta_{ C \to C'}$ whenever the system jumps from $ C$ to $C'$ along a spatio-temporal trajectory.
For the dynamical activity one has $\theta_{C \to C'}=1$ for all $C$ and $C'$ ($C \neq C'$) while for the entropy
production the increment will be $\theta_{C \to C'}=\ln (\omega_{C \to C'}/\omega_{C' \to C})$~\cite{LS99}.
We have already defined the increments for a global reaction-diffusion current in section~\ref{III}.  These increments, as 
we saw, affect the non-diagonal elements of the modified Hamiltonian. In contrast, for those time-dependent observables 
which are defined along a spatio-temporal trajectory and depend on the visited microstates, only the diagonal
elements of the modified Hamiltonian, depending on the observable, are changed. 

In either of these two cases we start with constructing the modified Hamiltonian $\hat{\cal H}(s)$ for the observable
under investigation. Let us denote the sum of the matrix elements of $i$th column of ${\hat h}(s)$ as $h_i$ for $i=1,\cdots,4$. 
For $\hat{\cal R}(s)$ and $\hat{\cal L}(s)$ they will be denoted by $r_i$ and $l_i$ respectively for $i=1,2$.  It can be shown 
that the summation vector $\langle \mathbb{1}|$ is the left eigenvector of the modified Hamiltonian given that
\begin{equation}
\label{GConstraint}
h_3-h_1=h_1-h_2=l_1-l_2=r_2-r_1, \;\; h_4=h_1\;.
\end{equation}
At the same time the eigenvalue of the modified Hamiltonian associated with that left eigenvector is given by
\begin{equation}
\Lambda^{\ast}(s)=l_1+r_1+(L-1)h_1
\end{equation}
in which $L$ is the size of the lattice. The equations~(\ref{GConstraint}) determine the value(s) of the conjugated 
field $s_0$ and also the probable constraints on the microscopic reaction rates under which the original and the effective
dynamics are equivalent in the sense of what was explained in section~\ref{III}.
\section{Concluding Remarks}
\label{VI}
In order to investigate the dynamics of a generic stochastic Markov process conditioned on an atypical value of an integrated current
during its time-translational invariance regime, one can modify its stochastic generator to build an effective (or auxiliary) stochastic 
generator for which the typical value of the integrated current in the steady state is equal to the atypical value of the integrated current 
in the original process. However, one realizes that the resulting effective process might be unphysical in the sense that it might contain    
non-local transitions.  In this paper we have shown that under some constraints on the microscopic reaction rates, the stochastic generator 
of the effective stochastic process can posses exactly the same dynamical rules as the original process does, at least for a specific value of 
the current under investigation. We have also shown that, depending on the process, this current might be entropic or non-entropic. Possible 
generalizations have also been discussed. Our approach might not be the only possible way to 
construct such effective stochastic process who shares identical features with the original stochastic process. It would be of great interest 
if one could find the general conditions under which the effective process would be physical in the sense that it only contains local transitions. 
On the other hand, we only considered the reaction-diffusion processes with nearest-neighbor interactions on open lattices. It would be 
interesting to investigate the processes with long-range interactions, not only on an open lattice but under periodic boundary conditions.   

\appendix*
\section{Derivation of the particle current formula~(\ref{jk})}
The time evolution of the average local particle density $\langle \rho_k \rangle (t)$ at the lattice site $k$ at time $t$ is given by
\begin{eqnarray}
\label{rho}
\frac{d}{dt} \langle \rho_k \rangle =J^R_{k-1 \rightarrow k}+J^R_{k \leftarrow k+1}
+J^D_{k-1 , k}-J^D_{k , k+1}
\end{eqnarray}
for $k=1,2,\dots ,L$ where $J^R_{k-1 \rightarrow k}$ and $J^R_{k \leftarrow k+1}$ are the average input current into the lattice site $k$, in the result of
reaction with the lattice sites $k-1$ and $k+1$ respectively. $J^D_{k , k+1}$ and $J^D_{k-1 , k}$ are also the net average 
diffusion current from the lattice site $k$ to $k+1$  and from $k-1$ to $k$ respectively. Note that $J^R_{0\leftrightarrows 1}=J^R_{L\leftrightarrows L+1}=0$
while $J^D_{0,1}=J^D_{L, L+1}\neq0$ which give the particle exchange with particle reservoirs at the boundaries. These quantities are given by 
\begin{eqnarray}
\label{R1}
J^R_{k-1 \rightarrow k} 
&=& \Big[ (\omega_{21}+\omega_{41}) \langle (1-\rho_{k-1})(1-\rho_{k}) \rangle \nonumber \\
&-& \omega_{12} \langle (1-\rho_{k-1})\rho_{k} \rangle 
+ \omega_{43} \langle \rho_{k-1}(1-\rho_{k}) \rangle \nonumber  \\
&-& (\omega_{14}+\omega_{34}) \langle \rho_{k-1}\rho_{k} \rangle \Big](1-\delta_{k,1})\\ 
\label{R2}
J^R_{k \leftarrow k+1}
&=& \Big[ (\omega_{31}+\omega_{41}) \langle (1-\rho_{k})(1-\rho_{k+1}) \rangle \nonumber  \\
&+& \omega_{42} \langle (1-\rho_{k})\rho_{k+1} \rangle 
- \omega_{13} \langle \rho_{k}(1-\rho_{k+1}) \rangle \nonumber  \\
&-& (\omega_{14}+\omega_{24}) \langle \rho_{k}\rho_{k+1} \rangle \Big](1-\delta_{k,L}) 
\end{eqnarray}
for $k=1,2,\dots ,L$ and 
\begin{eqnarray}
\label{D}
J^D_{k , k+1}
&=&\omega_{23} \langle \rho_{k}(1-\rho_{k+1})\rangle (1-\delta_{k,L}) (1-\delta_{k,0}) \nonumber  \\
&-&\omega_{32} \langle (1-\rho_{k})\rho_{k+1}\rangle (1-\delta_{k,L})(1-\delta_{k,0}) \nonumber  \\
&+& (\beta \langle \rho_{k}\rangle -\delta \langle 1-\rho_{k}\rangle)\delta_{k,L}  \nonumber  \\ 
&+& (\alpha \langle 1-\rho_{k+1}\rangle -\gamma \langle \rho_{k+1}\rangle)\delta_{k,0}  
\end{eqnarray}
for $k=0,\cdots,L$. The average local density of particles is related to the average particle current through the following
continuity equation~(\ref{ce}). Comparing~(\ref{rho}) and~(\ref{ce}) one finds the following relations for the average particle current 
$\langle j_{k} \rangle$ and the source term $S_{k}$ 
\begin{eqnarray*}
&& \langle j_{k} \rangle  =  J^R_{k \rightarrow k+1}-J^R_{k \leftarrow k+1}+J^D_{k , k+1} \; , \\
&& S_{k}  =  J^R_{k-1 \leftarrow k}+J^R_{k \rightarrow k+1} \; .
\end{eqnarray*}


\end{document}